\documentclass[copyright,creativecommons]{eptcs}
\providecommand{\event}{CCA 2010} % Name of the event you are submitting to
\usepackage{breakurl}             % Not needed if you use pdflatex only.

\title{A Local to Global Principle for the Complexity\\ of Riemann Mappings\\[0.3cm] \normalsize -- Extended Abstract --}
\author{Robert Rettinger
\institute{Department of Computer Science\\ FernUniversit\"at Hagen\\D-58084 Hagen, Germany}
\email{robert.rettinger@fernuni-hagen.de}}
\def\titlerunning{Complexity of Riemann Mappings}
\def\authorrunning{R. Rettinger}

\usepackage{graphicx}
\usepackage{amsfonts}

\usepackage{amssymb}
\usepackage{amsmath}
\usepackage{latexsym}

\def \IR {\mathbb{R}}
\def \IN {\mathbb{N}}

\def \ID {\mathbb{D}}
\def \IC {\mathbb{C}}
\def \IY {\mathbb{Y}}
\def \IG {\mathbb{G}}
\def \IB {\mathbb{B}}
\def \IH {\mathbb{H}}

\newcommand{\dom}{\mbox{dom}}

\newtheorem{definition}{Definition}
\newtheorem{lemma}{Lemma}
\newtheorem{theorem}{Theorem}
\newtheorem{claim}{Claim}
\newtheorem{corollary}{Corollary}

\begin{document}
\maketitle
\begin{abstract}
We show that the computational complexity of Riemann mappings can be bounded by the complexity needed to
compute conformal mappings locally at boundary points. As a consequence we get first formally proven
upper bounds for Schwarz-Christoffel mappings and, more generally, Riemann mappings of domains
with piecewise analytic boundaries.
\end{abstract}

\section{Introduction}

The Riemann mapping theorem is probably one of the most fundamental theorems in complex analysis.
Thus, not surprisingly, effectivization of this theorem was on the agenda ever since the theorem was proven.
The first general constructive solution for Riemann mappings were given by P. Koebe in 1910 using the osculation method
(see \cite{Hen86}). More constructivity results were added over the years based on different methods like
potential theoretic methods, circle packings and random walks. Using the osculation method, P. Hertling \cite{Her99}\
gave the first exact characterisation of the effectiveness of the Riemann mapping based on type-2-theory of effectivity.
Similar results were recently given for multiply connected domains (see \cite{MA09}).

Beside these effectiveness results it is known that the functor which maps simply connected domains to its Riemann mapping
is $\sharp P$-hard (see e.g. \cite{BBY07}). Even for the Riemann mapping of a single domain, hardness results are
known (see \cite{Re09}). On the other hand most natural domains seems to have further properties such as piecewise analyticity which seems to simplify
the task of computing Riemann mappings of these domains. A prominent example of such a restricted class of domains are polygons
and there exist seemingly fast algorithms to compute the Riemann mappings which
are also called Schwarz-Christoffel mappings in these cases. A strict proof that theses algorithms are polynomial time bounded, however, is not
known. Even worse, beside a few examples of domains for which the Riemann mapping is given explicitely, there is only one more general class known
to have polynomial time computable Riemann mappings: Using Bergmann kernels one can show that domains with analytic boundaries can indeed be
computed in polynomial time (see \cite{Re09}). But even for polygons or domains with piecewise analytic domains this method breaks down and leads to
exponential time bounded algorithms. Thus for these seemingly simple domains, the most efficient algorithms known so far are based on the random
walk technique (see e.g. \cite{BBY07}) and need exponential time.

In this paper we will present a more general class of domains with polynomial time computable Riemann mappings.
Our results are based on a local to global principle. That is, we show, that if locally polynomial time computable Riemann mappings or, more generally,
solutions to Dirichlet problems exist
then also the Riemann mapping of the domain is polynomial time computable.
As an immediate consequence of this result we get polynomial time algorithms for polygons and, more generally,
domains with piecewise analytic boundaries.

To prove our main result we will use the well known connection between Riemann mappings and harmonic functions (see Section 2 below).
Though there are well established solution methods for boundary value problems which work well in practice, it seems
to be very hard to prove that these methods work indeed in polynomial time. The most promising method is probably the finite element method.
Usually this method is using a more or less regular division of the domain, so that each of the subdomains (grid or triangulation) has roughly
the size of the precision one wants to achieve. This clearly cannot give a polynomial time bounded algorithm. But even if we adopt this method
and adjust the size of the subdomains similarly to the "layouts" which we will introduce in Section 3, there still is a problem which seems
hard to overcome: To bound the time needed to solve the induced linear system one has to get some bound on the conditioning, i.e. absolute values of the determinants of the underlying linear systems. Although one knows that in the end everything will work out nicely there is no general method known to calculate this "end". Whereas
such methods seems to be reachable for e.g. polygons, it is rather unlikely that we can compute bounds if we consider approximations to the
boundary which can vary with the precision one wants to achieve. We will use a technique similar to the finite element method known in scientific computation, but we will neither compute a linear
system explicitely nor compute any bounds on the conditioning. Instead we will compute harmonic functions on suitable subdivisions and update
these functions in several iterations. To analyse the time complexity of our method we will use a random walk technique rather than bounding
determinants.

We will end this section with a very short introduction to type-2-theory of effectivity. Afterwards in Section 2 we will prepare our main result
which we will present in Section 3. Finally, in Section 4 we will discuss some consequences of the main theorem.

Though we assume that the reader is familiar with the basic notions of type-2-theory as can be found e.g. in the textbook \cite{Wei00}, we have to adopt
the model slightly to get a natural notion of complexity. To this end let $\Sigma$ be a finite alphabet and $\Sigma^\ast$, $\Sigma^{\ast\ast}$ denote
the free monoid over $\Sigma$ and the class of functions $\Sigma^\ast\rightarrow\Sigma^\ast$, respectively. We say that
a function $f:\subseteq\Sigma^{\ast\ast}\rightarrow \Sigma^{\ast\ast}$ can be computed in time $t$ iff there
exists an oracle Turing machine $M$ which computes $f_M^{\sigma}(u)=v$ in at most $t(|u|)$ steps
whenever $f(\sigma)(u)=v$, i.e. the input of $f$ is used as an oracle. Combining this with classical complexity and standard tupling functions we can extend this definition to any products of $\Sigma^\ast$ and $\Sigma^{\ast\ast}$. To define complexity on more interesting classes we will use representations and realizations:
here a representation is a surjective functions $\nu:\subseteq\Sigma^\ast\rightarrow M$ or $\rho:\subseteq \Sigma^{\ast\ast}\rightarrow M$, where
$M$ is the class we want to represent. We will identify usual representations $f:\subseteq\IN\rightarrow M$ and
$g:\subseteq\Sigma^\omega\rightarrow M$ with representations $\nu$, $\rho$ so that $\nu(u)=f(|u|)$ and $\rho(\sigma)=g(\sigma(\varepsilon)\sigma(0)\sigma(00)...)$,
respectively, where we assume $0\in\Sigma$. A realization $F$ of a function $f:\subseteq M_1\rightarrow M_2$ between two represented classes
$M_1$, $M_2$ with representations $\delta_1$ and $\delta_2$, respectively, is a function $F:\dom(\delta_1)\rightarrow \dom(\delta_2)$ so that $\delta_2\circ F = f\circ\delta_1$ on the domain $\dom(f\circ\delta_1)$ of $f\circ\delta_1$. We say that $f$ is computable in time $t$ if there exists
a realizer of $F$ which is computable in time $t$.

To simplify things further, we will next introduce standard representations and use these representations later on without further mentioning.
That is, we call a function computable in time $t$ iff it is computable in time $t$ with respect to the standard representations.

We use the usual standard representations of the natural and dyadic numbers. Furthermore we use the signed digit representation for
the reals. By identifying the complex numbers with pairs of reals we get immediately standard representations for $\IN[i]$, $\IY[i]$ and $\IC$, respectively, where $\IY$ denotes the class of dyadics. Similarly, by identifying open balls $\ID_r(z)=\{ z'\in\IC\mid |z-z'|<r\}$ with the tupel $(r,z)$ we get a standard representation of the class $\IB$ of such balls. By combinding this with a standard representation
of the class of finite subsets of $\IN$ we get
our standard representation of the class $\IB^\ast$ of finite tuples of such balls. Finally, we will need a suitable representation of harmonic functions.
As we consider only a very restricted class of functions, i.e. all functions map some disc to $[0;1]$, the following representation omits
any information on the modulus of continuity, which can be computed easily. For $\rho\in\Sigma^{\ast\ast}$ and $u\in\Sigma^\ast$ let
$\rho\|_u$ denote the function $\rho\|_u:\Sigma^\ast\rightarrow\Sigma^\ast$ with $\rho\|_u(v)=\rho(uv)$. Then a name $\omega\in\Sigma^{\ast\ast}$
is mapped to a harmonic function $f$ under our standard representation iff $\omega\|_u$ is a name of $f(z)$ whenever $u$ is a name of
the dyadic $z$. Similar to balls we get a standard notation $\IH^\ast$ of all finite sequences of harmonic functions and finite sequences of
other represented objects. Beside we will need
piecewise harmonic functions. To simplify things we will in the sequel define for every
$s=((f_0,B_0),(f_1,B_1),...,(f_m,B_m))\in (\IH\times\IB)^\ast$ a unique function $h_s:\bigcup_i B_i\rightarrow [0;1]$ by
$h_s(z):=f_i(z)$ iff $z\in B_i\setminus\bigcup_{j<i} B_j$.

Beside the time needed to compute a certain function we will have to be careful about another resource: The lookahead of a machine on some input $\sigma$
is the function $l_M(\sigma):\IN\rightarrow\IN$ which gives for given $n\in\IN$ the maximum length of the question asked by the machine to $\sigma$ on
inputs of length $n$. We say that $M$ has linear lookahead iff there exists a konstant $k$ so that $l_M(\sigma)(n)\leq k\cdot n$ for all $n$ and
$\sigma$. This resource is especially important for iterations of functions. Assume that we want to iterate a polynomial number of times a polynomial
time computable function. If we are not carefull about the lookahead this can easily lead to functions which are no longer polynomial time computable.
If on the other hand we can guarantee that all these functions have linear lookahead with a common constant then the function given by this iteration
process is again polynomial time computable.

\section{Harmonic Functions, Dirichlet Problems and Riemann Mappings}

To simplify things we will in the sequel consider simply connected domains $G$ which are bounded by a Jordan curve $J$ so that
$\ID_{3/5}\subseteq G\subseteq \ID_{4/5}$. Any simply connected domain fulfilling the condition given in \cite{Her99}\ can be
transformed to such a domain by computable conformal mappings and, if the domain is bounded, even by dyadic polynomials.
Let the class of these restricted domains be denoted by $\IG$.

For $G\in\IG$ the classical Riemann mapping theorem states that
for any $z\in G$ there exists a unique conformal mapping $f(z):G\rightarrow \ID_1(0)$ so that $f(z)=0$ and $f'(z)>0$.
Let us denote this map by $f_z^G$ for the moment.
Obviously, any $z$ so that $f_z^G$ can be computed in polynomial time on some neighborhood of $0$ is itself polynomial time computable.
Furthermore, by using standard methods one can show that for $z$ and $z'$ in $G$, which are given via the
standard representation of $\IC$, we can compute $f_z^G\circ (f_{z'}^G)^{-1}$ in polynomial time. Thus we can restrict ourselves to
$z=0$ and we will write $f^G$ instead of $f_0^G$ for short.

It is not hard to give examples of domains $G\in\IG$ so that $f_z^G$ cannot be computed
in polynomial time throughout $G$ because the modulus of continuity of this function tends to infinity for these examples when we approach
the boundary of the domain. Thus, a polynomial time algorithm can only exist iff we compute $f_z^G$ only at values of a suitable distance to the
boundary of $G$. Analyzing the mentioned examples shows that possible upper bounds on the modulus of continuity are given by its non-euclidean
distance to $z$. To simplify things we will translate this to the euclidean distance $d(\cdot,\cdot)$ and introduce the following notion of
efficient computability inside $G$:

\begin{definition}
We say that a function $f$  is polynomial time computable inside $G$ iff there exists a machine $M$ and a polynomial $p$ so that
for all $n\in\IN$ and $z\in G$ so that $d(z,\partial G)>2^{-n}$, $M$ computs $f(z)$ up to precision $2^{-n}$ in time $p(n)$.
\end{definition}

Notice that polynomial time computability on $G$ does always imply polynomial time computability inside $G$ for functions defined on $G$.
Obviously, the converse does not hold.

Using the above notion we can formulate the well established relation between polynomial time computable Riemann mappings
and the solutions of Dirichlet problems (see \cite{BBY07}) as follows:

\begin{lemma}\label{lemma1}
Let $G\in \IG$ and $F:G\rightarrow G$ be the function defined by $F(z,w)= h_z(w)$ for all $z,w\in G$, where $h_z$ denotes the harmonic function
$h_z:\overline{G}\setminus \ID_{e^{-2n}}(z)\rightarrow [0;1]$ so that $h_z(v)=0$ for all $v\in\partial G$ and $h_z(v)=1$ for all
$v\in \partial \ID_{e^{-2n}}(z)$. If $F$ is polynomial time computable inside $G$ then the Riemann mapping $f^G$ is
polynomial time computable inside $G$.
\end{lemma}

Now let $G\subseteq \ID$ be a connected domain with regular boundary and $\varphi: \partial G\rightarrow [0;1]$
be a continuous function on the boundary of $G$. Notice that indeed the domains we need in Lemma \ref{lemma1}\ have regular boundaries.
Thus there exists a unique solution to the Dirichlet problem with boundary values given by $\varphi$. We will denote
this solution by $g_\varphi^G$. It is well known that the Dirichlet problem can be solved easily on the unit disk $\ID:=\ID_1(0)$
via the Poisson kernel function $P(w,z)=1/2\pi\cdot (1-|z|^2)/(|w-z|^2)$. This function is harmonic on $\IC\setminus\{ z\}$
for fixed $z$ and furthermore we have that $\int_{\partial\ID}\varphi(w)P(w,z)dw$ is a solution to the Dirichlet problem with boundary
values given by $\varphi$. We will see in the next section that this solution can be efficiently computed if $\varphi$ fulfills certain
regularity conditions. The above integral can be generalized to arbitrary Jordan domains $G$ by using the harmonic measure
$\omega_G(z)$, i.e. we have that $g_\varphi^G(z)=\int_{\partial G}\varphi(x)d\omega_G(z)(x)$ for all $z\in G$.

In our algorithm we will immitate the classical construction of solutions for the Dirichlet problem via subharmonic functions:
Using the regularity of the boundary we can find for each boundary point a local barrier and thus a neighborhood of the
boundary as well as a harmonic function on the common points of $\overline{G}$ and this neighborhood which coincede with $\varphi$ on
the boundary and is $0$ on all boundary points (of the neighborhood) which do not belong to the boundary of $G$.
Fixing all other points in $G$ to
be $0$ one repeatedly modifies this function by taking some disk inside $G$ and
replacing the value of the function inside this disk by the values given by the integral $\int_{\partial\ID}h(w)P(w,z)dw$ where $h$ denotes
the values of the unmodified function on the boundary of the disk. Iterating this modification with different disks will in the end
lead to the solution of the Dirichlet problem. We will need the following upper bound motivated by the above idea.

\begin{lemma}\label{lemma2}
Let $G\subseteq\ID$ be a domain with regular boundary and $\varphi:\partial G\rightarrow [0;1]$ be continuous. Then
we have for all Jordan domains $J\subseteq G$ and all intergrable functions $\Psi:\overline{G}\rightarrow [0;1]$ with
$\Psi\leq g_\varphi^G$:
\[ \int_{\partial J} \Psi(x) d\omega_J(z)(x) \leq g_\varphi^G(z)\]
for all $z\in J$.
\end{lemma}

The above idea motivated also the use of random walks to solve Dirichlet problems. This is an adopted version of Kakutanis
theorem to our situation:

\begin{theorem}\label{theo1}
Let $G\subseteq\ID$ be a domain, $\varphi$ be a continuous function on $\partial G$ and
let $R$ be a continuous function $R:G\rightarrow\IR$ with $1/3\cdot d(z,\partial G)<R(z)<2/3\cdot d(z,\partial G)$ for all $z\in G$. Furthermore let
the random process $x_t(z)$ be defined by
\[ x_0(z)=z,\quad x_t(z)=x_{t-1}(z)+e^{2\pi i \theta_t}\cdot R(x_{t-1}(z))\]
where $\theta_0, \theta_1, ...$ is a sequence of independent random variables, uniformly distributed in $[0;1]$.

Then with probability 1, $lim_{t\to\infty}x_t(z)$ exists, $lim_{t\to\infty}x_t(z)\in\partial G$
 and $E(\varphi(lim_{t\to\infty}x_t(z))=g_\varphi^G(z)$.
\end{theorem}

Discretizing this idea leads to the random walk algorithm in \cite{BBY07}. To get the result to precision $2^{-n}$ one has in each
 $x_t$ to take an exponential number of $\theta_i$. Thus clearly this cannot give polynomial time algorithms. On the
other hand, however, the number $t$ of iterations can be bounded by a polynomial to get precision $2^{-n}$.
We give this result by a modification of the above theorem.

\begin{lemma}
Let $G$, $\varphi$ and $R$ be given as in the above theorem. Furthermore let $|\varphi(x)-\varphi(y)|<k_1 |x-y|^{2/3}$ for all $x,y$ in $\partial G$ and some
constant $k_1$. If we define for $z\in G$ the random process $x_t$ by
\[ x_0(z)=z,\quad x_t(z)=\left\{\begin{array}{ll} z' & \mbox{ if } d(x_{t-1}(z),\partial G)<2^{-3n}\\
x_{t-1}(z)+e^{2\pi i \theta_t}\cdot R(x_{t-1}(z)) & \mbox{ otherwise}\end{array}\right.\]
where $z'$ is some boundary point near $x_{t-1}(z)$,  then
\[ |E(\varphi(x_T(z))-g_\varphi^G(z)|<2^{-2n}\]
where $T=k_2\cdot n^3$ for a suitable constant $k_2$ and $\varphi(y):=0$ for all $y\notin\partial G$.
\end{lemma}

The previous lemma will provide a lower bound by using the following lemma:

\begin{lemma}\label{lemma4}
Let $G$, $\varphi$, $R$ and $x_t$ be given as in Theorem \ref{theo1}. Furthermore let $t$ and $\varepsilon>0$ be given. Then
for any function $\Psi:G\rightarrow [0;1]$ with $\Psi(z)\geq E(\varphi(x_t(z)))-\varepsilon$ for all $z\in G$ we have
\[ \int_{\partial \ID_{R(z)}(z)} \Psi(w)P(w,z)dw \geq E(\varphi(x_{t+1}(z))) -\varepsilon \]
for all $z\in G$.
\end{lemma}

To get a starting point we thus have to find some neighborhood of the boundary of $G$ and a (sub-)harmonic function on
$\Psi_0$ on this neighborhood so that $\Psi_0(z)\geq E(\varphi(x_1(z)))$ for all $z$.
For the special domains which we will need by Lemma \ref{lemma1}\ this is nearly trivial: Let therefore for $G\in\IG$, $n\in\IN$ and some $z_0\in G$ with $d(z_0,\partial G)>2^{-n}$ by $G_{z_0}^n$ denote the domain $G_{z_0}^n=G\setminus \ID_{e^{-2n}}(z_0)$.
Furthermore let a covering $N_i$ of the boundary $\partial G$ be given so that locally the Dirichlet problem can be computed efficiently on each
of these neighborhoods. We will define this notion formally in the next section.
Then we can choose $\Psi_0$ near the boundary, i.e. on the $N_i$, by $\Psi_0\equiv 0$ and furthermore choose
$\Psi_0$ on $N_{z_0}^n:={G_{z_0}^n\cap \ID_{1/2\cdot 2^{-n}}(z_0)}$ to be the solution
$g_\varphi^{N_{z_0}^n}$ of the Dirichlet problem with boundary values given by $\varphi$ where
$\varphi(\partial\ID_{e^{-2n}}(z_0))=1$ and $\varphi(\partial\ID_{1/2\cdot 2^{-n}}(z_0))=0$. Outside these sets we let $\Psi_0$ be $\Psi_0\equiv 0$.
Then we compute the sequence $\Psi_0$, $\Psi_1$, ...,
$\Psi_{k_2\cdot n^3}$ where $\Psi_{t+1}$ is computed from $\Psi_{t}$ by
\[ \Psi_{t+1}(z):=\int_{\partial \ID_{R(z)}(z)} \Psi_t(w)P(w,z)dw \]
if $z\notin N_{z_0}^n\bigcup N_i$ and
\[ \Psi_{t+1}(z):=g_{\hat{\Psi_t}}^N(z)\]
if $z\in N$ for some $N=N_i$ or $N=N_{z_0}^n$. Here $R$ denotes again a function with
$1/3\cdot d(z,\partial G)<R(z)<2/3\cdot d(z,\partial G)$ for all $z\in G$ and $\hat{\Psi_t}$ is given by
$\hat{\Psi_t}\equiv 0$ on $\partial G$, $\hat{\Psi_t}\equiv 1$ on $\partial\ID_{1/2\cdot 2^{-n}}(z_0)$ and
$\hat{\Psi_t}=\Psi_t$ on all other points.

By Lemma 2, 3 and 4 $\Psi_{k_2\cdot n^3}$ is then an acceptable approximation of the corresponding Dirichlet problem which we will need to apply Lemma 1.

\section{The Main Theorem}

In this section we will show that the ideas developed in the last section lead indeed to a polynomial time algorithm
for the Riemann mapping. Probably the easiest task of this proof is the characterization of those $G\in \IG$ which
allow to compute the function $R$ of Theorem \ref{theo1}\ effectively. To this end we have to compute
the distance function to the boundary up to constant factors. Here the factor is somewhat arbitrary.
We will call this class of domains polynomial time computable. Despite the fact that we have motivated
the following definition by the ideas given in the previous
section, the definition itself is quite natural and equivalent definitions have been used before to characterize
computably efficient domains (see e.g. \cite{Re03}).

\begin{definition}
We call $G\in \IG$ polynomial time computable iff there exists a polynomial time computable function $F:G\rightarrow [0;1]$ so that
for all $z\in G$ we have $(1-1/4)\cdot d(z,\partial G)<F(z)<(1+1/4)\cdot d(z,\partial G)$.
\end{definition}

Let for the time beeing $G\in\IG$ be polynomial time computable and $F$ its approximation function of the previous definition.
Then the function $R:G\rightarrow [0;1]$ with $R(z):=1/2\cdot F(z)$ for all $z\in G$ is also polynomial time computable and we have indeed
\[ 1/3\cdot d(z,\partial G)<R(z)<2/3\cdot d(z,\partial G)
\] for all $z\in G$. We even have more: Let $r(z):=1/4\cdot F(z)$. Then for every $x\in \ID_{r(z)}(z)$ we have
\[ 1/3\cdot d(x,\partial G)<R(z)<2/3\cdot d(x,\partial G).\]

We will next discuss the computation of the integral $\int_{\partial \ID_{R(z)}(z)} \Psi_t(w)P(w,z)dw$. It is well known that
integration leads easily to $\sharp$-P hard problems and thus there is no polynomial time upper bound known. However, if
we restrict ourselves to harmonic or analytic functions integration can be done in polynomial time (see e.g. \cite{Mu87}).
The following result can be easily proven by such techniques. Let for a given ball $B=\ID_r(z)$ the ball $\ID_{2r}(z)$ be denoted
by $2B$.

\begin{lemma}\label{lemma5}
Let $G\in \IG$, $d>0$ be given. Furthermore assume that there exists a polynomial time computable function
$S:\IN\rightarrow (\IH\times\IB)^\ast$ so that with $S(n)=((f_0,B_0),...,(f_n,B_n))$ we have  $2B_i\subseteq G$ for all $i$ and
all functions $f_i:B_i\rightarrow [0;1]$ can be extended to harmonic functions on $2B_i$ for all $i$.

Then there exists a polynomially time
bounded machine $M$ with linear lookahead which computes for given $n\in\IN$, $z\in G$ and $r\in\IY$  the integral
\[\int_{\partial \ID_{r}(z)} h_{S(n)}(w)P(w,z)dw\]
whenever $\ID_r(z)$ is covered by the $B_i$ and $r>2^{-d\cdot n}$.
\end{lemma}

Finally, we have to formally define what we mean by "locally efficient solutions of Dirichlet problems". Roughly speaking,
this means that we can cover the boundary by neighborhoods so that for each of these neighborhoods and any piecewise harmonic
function on the boundaries of these neighborhoods one can compute the solution of the corresponding Dirichlet problem in polynomial
time.

\begin{definition}
An admissible neighborhood in a domain $G\subseteq \ID$ is a tuple $(z,r,D)$ where $z\in\partial G$, $r\in\IY$ and
$D\subseteq G$ is a simply connected domain so that $\ID_{5 r}(z)\cap G\subseteq D$ and $\ID_{5 r}(z)\cap \partial G$ is a simple, connected curve.
An admissible $n$-covering of such an admissible neighborhood is a finite sequence $(B_0,...,B_{m+1})\in\IB^\ast$
so that $2B_i\subseteq G$ for $i=1,...,m$, the diameter of $B_0$ and $B_{m+1}$ are at most $2^{-3\cdot n}$
and $\bigcup_i B_i\supseteq \partial D\setminus\partial G$.
Furthermore the core of an admissible neighborhood is the set $\ID_r(z)$.

Let $\varphi:G\rightarrow [0;1]$ be given. We say that $G\subseteq\ID$ admits locally polynomial time solutions of the Dirichlet problem with boundary values $\varphi$ iff there exist polynomial time computable functions $S:\IN\rightarrow (\ID\times \IY\times\IB^\ast)^\ast$
and $U:\partial\IN^2\times\IH^\ast\times \IC\rightarrow\IC$ so that for all $n$ there
exists a sequence $(z_0,r_0,D_0)$, $(z_1,r_1,D_1)$, ..., $(z_m,r_m,D_m)$ of admissible neighborhoods in $G$ so that
\begin{enumerate}
\item $S(n)=((z_0,r_0,C_0),...,(z_m,r_m,C_m))$,
\item the cores of these admissible neighborhoods cover $\partial G$,
\item the $C_i$ are admissible $n$-coverings of the $(z_i,r_i,D_i)$ and
\item for all $i\in\IN$ and $f=(f_0,...,f_{s+1})\in \IH^\ast$ we have that $g_{\hat{\varphi}}^{D_i}=U(n,i,f,\cdot)$
whenever $C_i=(B_0,...,B_{s+1})$, $\hat{\varphi}=\varphi$ on $\partial G$ and $\hat{\varphi}=h_{((f_0,B_0),...,(f_{s+1},B_{s+1}))}$
on $\partial D_i\setminus \partial G$.
\end{enumerate}
$S$ and $U$ are then called solutions of the problem. The balls in $C_i$ are called boundary nodes of the solution.
\end{definition}

With this preparations we can nearly implement the idea at the end of the previous section literally. Of course we can neither keep the
values of $\Psi_t(z)$ for all $z$ nor can we update $\Psi_t(z)$ to $\Psi_t(z)$ for each $z\in G$ without any assumptions on $\Psi_t$. We
will construct $\Psi_t$ to be piecewise harmonic. Thus we can indeed keep track of any of the values of $\Psi_t$ easily. To this end,
however, we need some partition of $G$ so that all $\Psi_t$ are harmonic on each part of this partition. Furthermore there should not
too many of these parts. We will call such partitions "layouts". The following definition will be used inside the proof of
our main theorem.

\begin{definition}
Let $n\in\IN$ be given, $\varphi:G\rightarrow [0;1]$ and $U$, $S$ be locally polynomial time solutions of the Dirichlet problem with
boundary values $\varphi$. Furthermore let $S(n)=((z_0,r_0,C_0),...,(z_m,r_m,C_m))$.

We say that $(B_0,...,B_s)\in\IB^\ast$ is an $n$-layout for this solution iff $G\setminus \bigcup_i \ID_{3\cdot r_i}\subseteq \bigcup_j B_j$
and for each $B_j$ we have that $2B_j\subseteq G$. $s$ is called the size of the $n$-layout.

Now let $G$ be polynomial time computable. We say that $G$ admits polynomial time computable layouts iff there exists a polynomial time
computable function $V:\IN\rightarrow\IB^\ast$ so that $V(n)$ is an $n$-layout and for each $\ID_r(z)$ in $V(n)$ we have
$r=r(z)=1/4 F(z)$.
\end{definition}

The last step we need to prove our main theorem is a proof that polynomial time computable layouts exists. We will give the main idea
of such a proof. A formal proof will be given in the full version of this paper.

\begin{lemma}\label{lemma6}
Let $G\subseteq\ID$ be polynomial time computable. Furthermore let $\varphi:G\rightarrow[0;1]$ be given so that $G$ admits
locally polynomial time solutions $U$ and $S$ of the Dirichlet problem with boundary values $\varphi$. Then $G$ admits also
polynomial time layouts for this solution.
\end{lemma}

\noindent {\bf Proof (Idea):}

First notice, that the construction of layouts is simple. We can start with a point $z$ inside $G$, compute the correct radius of the
corresponding disc, i.e. $1/4\cdot F(z)$, and continue with some point on the boundary of this disc, while checking wether $G$ is already
covered by the neighborhoods given by $U$ and the constructed discs. It is thus not hard to prove the following claim:

\begin{claim}
$G$ admits polynomial time layouts for the solution iff there exists a polynomial $p$ so that for each $n$ there exists an $n$-layout
of size $p(n)$ for the solution.
\end{claim}

To show that such $n$-solutions of polynomial size always exist we proceed as follows:
Let $K_0$ be some connected component of $G\setminus\bigcup_i 3B_i$. where the $B_i$'s are the cores of the corresponding admissible neighborhoods.
W.l.o.g. we can assume that there exists some $d>0$ (common for all $n$) so that all boundary nodes and all $B_i$'s have diameter
at least $2^{-d n}$. Now let $z\in \partial K$ be given so that $d(z,\bigcup_i B_i)$ is minimal and add the disc $\ID_{r(z)}(z)$ to the layout. Furthermore
mark the core $B_i$ by level $0$. Next proceed with $K_0$ replaced by $K_1=K\setminus \ID_{2r(z)}(z)$ and $\bigcup_i B_i$ replaced
by $\ID_{r(z)}(z)\cup \bigcup_i B_i$. Notice, that any in this way added disc has marked some core. Generally, if $j$ is minimal so that
$z$ belongs to $\partial K_j$ we mark the corresponding core by level $j$.

It is not hard to show that each core can be marked at most $6\pi$ times at each level. Furthermore the maximum level by which a core can be marked
is bounded by $\log_{1.8} 2^{d\cdot n}$ because if we go from level $k$ to $k+1$ the diameters of the considered discs grow at least by a factor
$(1+\sqrt{3/4})$ as can be seen by elementary geometry.

Thus in the end we get a layout of size $c\cdot m\cdot n$ for a suitable constant $c$ (which does not depend on $n$) where $m$ denotes
the number of cores. As this number is polynomially bounded by the definition of locally polynomial time solutions of Dirichlet problems,
we get altogether a polynomial upper bound on the number of discs in the layout.\\ \mbox{} \hspace{14cm} $\Box$\\[1cm]

Now our main result follows easily by the idea given at the end of the previous section:

\begin{theorem}\label{main}
Let $G\in\IG$ be a polynomial time computable simply connected domain which admits
locally polynomial time solutions of the Dirichlet problem with boundary values $\equiv 0$.
Then, for any $z\in G$, the corresponding Riemann mapping $f_z^G$ is polynomial time computable inside $G$.
\end{theorem}

\noindent {\bf Proof (Idea):}

By Lemma \ref{lemma1}\ the theorem is proven once we have shown that the Dirichlet problem on $G_{z_0}^n$ with
boundary values $\equiv 0$ and $\equiv 1$ can be proven in time $p(n)$ where $p$ is a suitable polynomial.

First notice that $G$ admits locally polynomial time solutions of the Dirichlet problem iff $G_{z_0}^n$ admits
such solutions because the boundary componend $\ID_{e^{-2n}}(z_0)$ admits clearly a constant number of admissible neighborhoods and
corresponding solutions. By Lemma \ref{lemma6}\ we can therefore compute a polynomial time layout for this solution.

Let us call any of the boundary or inner nodes just nodes and let $B_i$ be the cores of the solution.
We start with exactly the $\Psi_0$ given at the end of the previous
section. Then we update every node $D=\ID_r(z)$ as follows: If $D\cap \bigcup_i 3B_i\not=\emptyset$ then we have by definition
of admissible neighborhoods that $2D\subseteq 5B_j$ for some $B_j$ and thus we compute the new values on $D$ by
computing the solution of the local Dirichlet problem on the corresponding
neighborhood. Otherwise we compute the new values on $D$ by integration (Lemma \ref{lemma5}).

Let us call the procedure of updating all nodes a round. Each round can be computed in polynomial time. Thus to proof our main theorem
it remains to show that we only need a polynomial number of rounds to approximate the solution of the Dirichlet problem to precision $2^{-n}$.
However, at the end of each round we get some piecewise harmonic function on $G$ and a simple inductive argument shows that
the piecewise harmonic function $f_t$ we get after $t$ rounds is larger than the function $\Psi_t$ which we have introduced at the
end of the previous section. Especially we get $f_t(z)>E(\varphi(x_t(z)))$ where $\varphi$ is the corresponding boundary value ($\equiv 0$ on $\partial G$ and $\equiv 1$ on the other boundary component). Thus we get that after $T=k_2\cdot n^3$ rounds we have that $f_T$ is larger than the solution minus
$2^{-2n}$. By Lemma \ref{lemma2}\ we have on the other hand that for any $t$ the function $f_t$ must be smaller than the solution. Thus
after $T$ rounds we get a correct approximation of the solution.

As the number of nodes is polynomially bounded and any update can be done in polynomial time,
we iterate polynomial times a polynomial time algorithm with linear lookahead. That is the overall algorithm is indeed polynomial time bounded.\\ \mbox{} \hspace{14cm} $\Box$\\[1cm]

The following well known result shows that it suffices to consider approximations of the boundary of $G$ in the above construction.
Thus the above result although holds iff locally polynomial time solutions of the Dirichlet problem exists for approximations of $G$, where
these approximations may depend on the precision one wants to achieve. We omit the details.

\begin{lemma}
Let $\gamma_1$ and $\gamma_2$ be two closed Jordan curves with $d(\gamma_1,\gamma_2)<2^{-3n}$. Furthermore let $h_1$ and $h_2$ be two
continuous functions on $\gamma_1$ and $\gamma_2$, respectively, with the following continuity property: if $x_1\in\gamma_1$ and $x_2\in\gamma_2$
and $|x_1-x_2|<2^{-2n}$ then $|u_1(x_1)-u_2(x_2)|<2^{-n}$. Then the solutions $u_{h_1}$ and $u_{h_2}$ of the corresponding Dirichlet problem fulfills
\[ |u_{h_1}(z)-u_{h_2}(z)|<2^{-n+1}\]
for all $z$ inside $\gamma_1$ and $\gamma_2$ with $d(z,\gamma_1)>2^{-n}$ and $d(z,\gamma_2)>2^{-n}$.
\end{lemma}

\section{Discussion}

We end this paper with a short list of applications  of and some remarks on our main theorem.

With our main theorem it is now easy to prove that Schwart-Christoffel mappings are polynomial time computable where we do not even
have to mention the parameter problem. To define admissible neighborhoods we have simply to use the fact that
mappings for triangles can be computed in polynomial time because here the parameter problem is trivial (see e.g. \cite{Hen86}).
Thus we can choose simply triangles as local neighborhoods and use the Schwarz-Christoffel mappings and the relection principle
to solve Dirichlet problems locally:

\begin{corollary}
Let $G$ be a polygon so that any vertex is polynomial time computable. Then for any $z\in G$
the Riemann mapping $f_z^G$ is polynomial time computable.
\end{corollary}

A similar idea shows that we can extend this result to piecewise analytic boundaries. Here we use the fact that we can map the
analytic boundary locally one to one onto some intervall. To get an admissible neighborhood we therefore just have to use this
map, take a small triangle and map it back by the inverse of the first mapping. Thus by applying again Schwarz-Christoffel mappings for triangles we
can solve locally Dirichlet problems.

\begin{corollary}
Let $G$ be a simply connected domain with piecewise analytic, polynomial time computable boundary. Then for any $z\in G$
the Riemann mapping $f_z^G$ is polynomial time computable.
\end{corollary}

Notice that this second result cannot be extended to domains where the boundary is approximated by piecewise analytic boundaries which may differ
for different precisions. To this end, more results on what kinds of approximation could lead to polynomial time algorithms would be interesting.

On the other hand such results can be easily shown for polygons where the number of vertices is bounded polynomially
by $n$ and the vertices can be computed uniformly in polynomial time.

Combining the previous result with the result \cite{Re09}\ we get an improved lower bound on the complexity of Riemann mappings
for single domains. Details can be found in the original paper.

\begin{corollary}
If $\sharp P\not=_{sel}P$ then there exists a polynomial time computable simply connected domain $G$ so that for any computable $z\in G$ we have
that $f_z$ is not even polynomial time computable on a compact subset of $G$.\\[2cm]
\end{corollary}

\bibliographystyle{eptcs} % or whatever you prefer

\end{document}